\documentclass[letter,tradiabstract]{aa}  

\usepackage{graphicx}
\usepackage{txfonts}
\usepackage{xcolor}%
\usepackage{threeparttable}
\usepackage{xr}
\usepackage{soul}
\usepackage[hidelinks,colorlinks=true,linkcolor=blue,citecolor=blue]{hyperref}
\usepackage{comment}
\newcommand{\CI}{{}C\,{\sc i}}

\newcommand{\kms}{\mbox{\textrm{km}\,s$^{-1}$}}

\newcommand{\skss}{Sk$-$67\,2}
\newcommand{\skse}{Sk$-$68\,137}
\newcommand{\skoft}{Sk\,143}

\DeclareDocumentCommand{\PN}{s m O{}}{{\color{purple}\IfBooleanTF{#1}{\color{gray} \setstcolor{purple}\st{#2} \color{purple}#3}{{\sl [PN: #2]}}}}

\begin{document} 
\newcommand{\ioffe}{Ioffe Institute, {Polyteknicheskaya ul. 26}, 194021 Saint-Petersburg, Russia \label{ioffe}}

\newcommand{\iap}{Institut d'Astrophysique de Paris, CNRS-SU, UMR\,7095, 98bis bd Arago, 75014 Paris, France \label{iap}}

   \titlerunning{CO absorptions in the Magellanic Clouds}

\title{First detections of CO absorption in the Magellanic Clouds and direct measurement of the CO-to-H$_2$ ratio}

   \author{S.~A.~Balashev\inst{\ref{ioffe}}
          \and
          D.~N.~Kosenko\inst{\ref{ioffe}}
          \and   
          P.~Noterdaeme\inst{\ref{iap}}
          }

   \institute{\ioffe -- 
              \email{s.balashev@gmail.com}
              \and \iap 
             }
   \date{}
   \date{Received November 06, 2024; accepted March 13, 2025}

\abstract{ Molecular hydrogen (H$_2$) is by far the most abundant molecule in the Universe. However, due to the low emissivity of H$_2$, carbon monoxide (CO) is widely used instead to trace molecular gas in galaxies. The relative abundances of these molecules is expected to depend on both physical (e.g., density) and chemical (e.g., metal enrichment) properties of the gas, making direct measurements in diverse environments crucial. We present a systematic search for CO in absorption toward 34 stars behind H$_2$ gas in the Magellanic Clouds using the Hubble Space Telescope. We report the first two definitive detections of CO absorption in the Large Magellanic Cloud (LMC) and one in the Small Magellanic Cloud (SMC), along with stringent upper limits for the remaining sightlines. Non-detections of CO are consistent with models of low thermal pressures and/or low metallicities while detections at the lower metallicities of the Magellanic Clouds require higher thermal pressures, $P_{\rm th}=10^5-10^6$~K\,cm$^{-3}$ than detections the Milky Way at similar $N({\rm H_2})$. Notably, the high density derived from the rotational excitation of CO towards SK\,143 in the SMC suggests full molecularization of CO in the absorbing cloud, with CO/H$_2 = 8.3^{+2.0}_{-1.6} \times 10^{-5}$ consistent with the standard ratio ($3.2\times10^{-4}$) measured in dense molecular gas in the Milky Way, scaled to the SMC’s 0.2\,$Z_{\odot}$ metallicity.}


   \maketitle
%

\section{Introduction}
\label{sec:intro}
The Small and Large Magellanic Clouds (SMC and LMC) are nearby dwarf galaxies, tidally interacting with the Milky Way (MW) and exhibiting lower average metallicities than the latter \citep{Russell1992,Kosenko2024}. The proximity of the Magellanic Clouds (MCs) has enabled extensive studies across the entire electromagnetic spectrum, from radio \citep[e.g.][]{Bruns2005} to $\gamma$-rays \citep[e.g.][]{Abdo2010, Abdo2010b}.
Notably, the molecular gas in the interstellar medium (ISM) of these galaxies has long been investigated through CO emission \citep[see review by][]{Fukui2010}. While emission line studies provide a global view of the molecular content, they are mostly sensitive to dense and warm molecular gas producing lines from e.g. HCO$^{+}$ and high rotational CO transitions. Moreover, these are limited by the resolution of the telescopes, which at the distance of the MCs \citep{Pietrzynski2019, Graczyk2020}, correspond to physical scales ranging from $\sim40$ pc for \mbox{NANTEN} \citep[e.g.][]{Mizuno2001, Fukui2008} to $\lesssim0.4$ pc for ALMA \citep[e.g.][]{Indebetouw2013, Jameson2018}. 

Absorption line studies towards bright, point-like sources such as stars or quasars offer a more detailed view of local gas properties. Not only these also probe low-excitation and diffuse gas, that is, including  "CO-dark" molecular gas that remains invisible in CO emission but also permit accurate, simultaneous measurements of the species column densities. 
 The angular sizes of these sources are also much smaller than typical ISM structures, virtually eliminating transverse spatial averaging. Although some averaging along the line of sight is inevitable, velocity decomposition in absorption line profiles alleviates this effect.

In fact, absorption studies laid the foundation for our discovery of the interstellar medium (ISM) and our basic understanding of its properties. The first detection of H$_2$ absorption in the ISM \citep[][]{Carruthers1970} was soon followed by that of CO \citep{Smith1971}, opening the way to the study of both molecules using larger samples of bright nearby stars \citep[e.g.][]{Savage1977,Federman1980}. 
A few decades later, the Hubble Space Telescope (HST) enabled more detailed studies of the physical properties of CO-bearing gas along Galactic sightlines \citep{Sheffer2007, Sheffer2008}, resolving its rotational population \citep{Sonnentrucker2007}. At the same time, large ground-based telescopes led to the detection of electronic CO absorption in intervening galaxies at $z\sim1.7-2.7$ towards background quasars \citep[][and possibly \citealt{Ma2015}]{Srianand2008,Noterdaeme2009,Noterdaeme2010,Noterdaeme2011,Noterdaeme2017,Noterdaeme2018}, and at $z\sim3$ in the host galaxy of a $\gamma$-ray burst \citep{Prochaska2009}. We also note the recent detection of intervening CO absorption in the radio domain at $z=0.05$ towards a $z=1.3$ quasar by \citet{Combes2019}.
CO has also been detected in absorption within circumnuclear regions in active galaxies through radio \citep{Emonts2024}, as well as IR observations \citep{Shirahata2013,Onishi2021,Ohyama2023}. 

It is hence somewhat paradoxical that possible CO absorption has only been reported along three sightlines through the Magellanic Clouds (MCs) to date: Sk$-$67\,5, Sk$-$68\,135, and Sk$-$69\,246 \citep{Bluhm2001,Andre2004}\footnote{CH, CH$^+$, C$_2$, { C$_3$ and CN} were firmly detected in other MC sightlines by \citet{Welty2006,Welty2013}.}. These claims were based solely on data around the CO C-X band at 1088\,\AA, and given the insufficient spectral resolution and known calibration issues with FUSE \citep{Kosenko2023}, these detections are debatable.
\cite{Welty2016} also mentioned the presence of CO absorptions towards Sk 143 and Sk-68 73 in HST/STIS and FUSE spectra, respectively, but did not provide further details.

Several, more convenient A-X bands are available at $\lambda\gtrsim1300$ \AA, accessible via the Hubble Space Telescope (HST). In this Letter, we present a systematic search for and study CO absorption in the Magellanic Clouds (MCs) based on archival HST data. We do not confirm the previous claims, but report the first definitive detections of CO absorption along two sightlines in the Large Magellanic Cloud (LMC) and one in the Small Magellanic Cloud (SMC). This allows us to directly derive and discuss the relative abundance of CO and H$_2$ for the first time in these environments.

\section{Data and analysis}
\label{sec:data}

In order to search for CO absorption, we scrutinised archival data from the Cosmic Origin Spectrograph (COS) and the Space Telescope Imaging Spectrograph (STIS) of UV-bright stars in the LMC and SMC, where H$_2$ has been detected \citep{Welty2012, Kosenko2023} with column densities $N({\rm H_2})\gtrsim 10^{19.5}$ (here and below all column densities are in cm$^{-2}$). 

For each sightline, we considered all available spectra during the analysis.
We constructed the local continuum over the expected positions of CO bands (both $^{12}$CO and $^{13}$CO) using an iterative B-spline model, constrained by unabsorbed regions. The continuum was then visually inspected and corrected if necessary. We used a  compilation of CO transitions from \citet{Dapra2016}, which includes A-X bands from $(0-0)$ to $(9-0)$, as well as the d-X 5-0 band.
We performed Voigt profile fitting of the data, along with Bayesian inference through Affine-invariant sampler \citep{Goodman2010} to obtain constraints (including upper limits) on the CO column densities and other model parameters. The number of components is based on the associated \CI\ absorption. The line profiles with line spread function which was chosen to be Gaussian for STIS spectra and provided by Space Telescope Science Institute for COS\footnote{see https://www.stsci.edu/hst/instrumentation/cos/performance/spectral-resolution}.

Due to the limited spectral resolution, the lines corresponding to transitions from CO rotational levels are either unresolved (for COS spectra) or barely resolved (for STIS spectra). To obtain consistent column densities across different rotational levels, we thus assumed homogeneous excitation of CO within the absorption gas: we used a single Doppler parameter for all CO rotational levels and tied the column densities of the various rotational levels using a one-zone excitation model. Details of the model are provided in Appendix~\ref{sect:CO_model}.

We accounted for uncertainties in continuum placement following \citet{Noterdaeme2021}. Briefly, for each CO band, we estimated the pixel dispersion in regions near the absorption line and applied a hierarchical Bayesian model, with a factor $h$ representing continuum variations.

We used Gaussian priors on the component velocities derived from the \CI\ absorption lines (except for sightlines towards \skss\ and \skoft, where we used CH lines detected in high-resolution optical spectra). For the Doppler parameters we conservatively assumed Gaussian prior with mean of 1.0 and standard deviation of 0.3 $\rm km\,s^{-1}$. For temperature (which sets the collisional rates) we used log normal priors $\log T_{\rm k} [\rm K]=1.7\pm0.3$ and $1.2\pm0.1$, expressing the mean $\pm$ standard deviation (std) for sightlines with CO non-detection and detection, respectively. This choice was motivated by observations \citep[see e.g.][]{Balashev2019, Klimenko2024} and \texttt{Meudon PDR} modelling (see Fig.~\ref{fig:temp_priors} and discussion in Appendix~\ref{sect:appendix_temp}). For sightlines with CO detections, we used a flat prior for the number densities $\log n$, while for those without CO detection, we applied a log-normal prior, $\log n [\rm cm^{-3}] = 2.5 \pm 0.3$, corresponding to typical values in molecular clouds \citep[e.g.][]{Draine2011, Klimenko2020}. For the continuum variation factor $h$ (unitless), we assumed a Gaussian prior with $h=1.0\pm0.3$.

All data analysis, including profile fitting, CO population calculations, and posterior distribution inference, was performed using the publicly available code {\tt spectro}\footnote{https://github.com/balashev/spectro}. For each parameter, we report both the point estimate and interval estimates, which correspond to the maximum a posteriori probability and the 0.683 highest posterior density interval, respectively. The latter represents the statistical uncertainty under the given model assumptions, which should be taken with caution.

\section{Results}
\label{sec:results}

\begin{figure*}[!ht]
\centering
\resizebox{1.0\hsize}{!}
        {\includegraphics{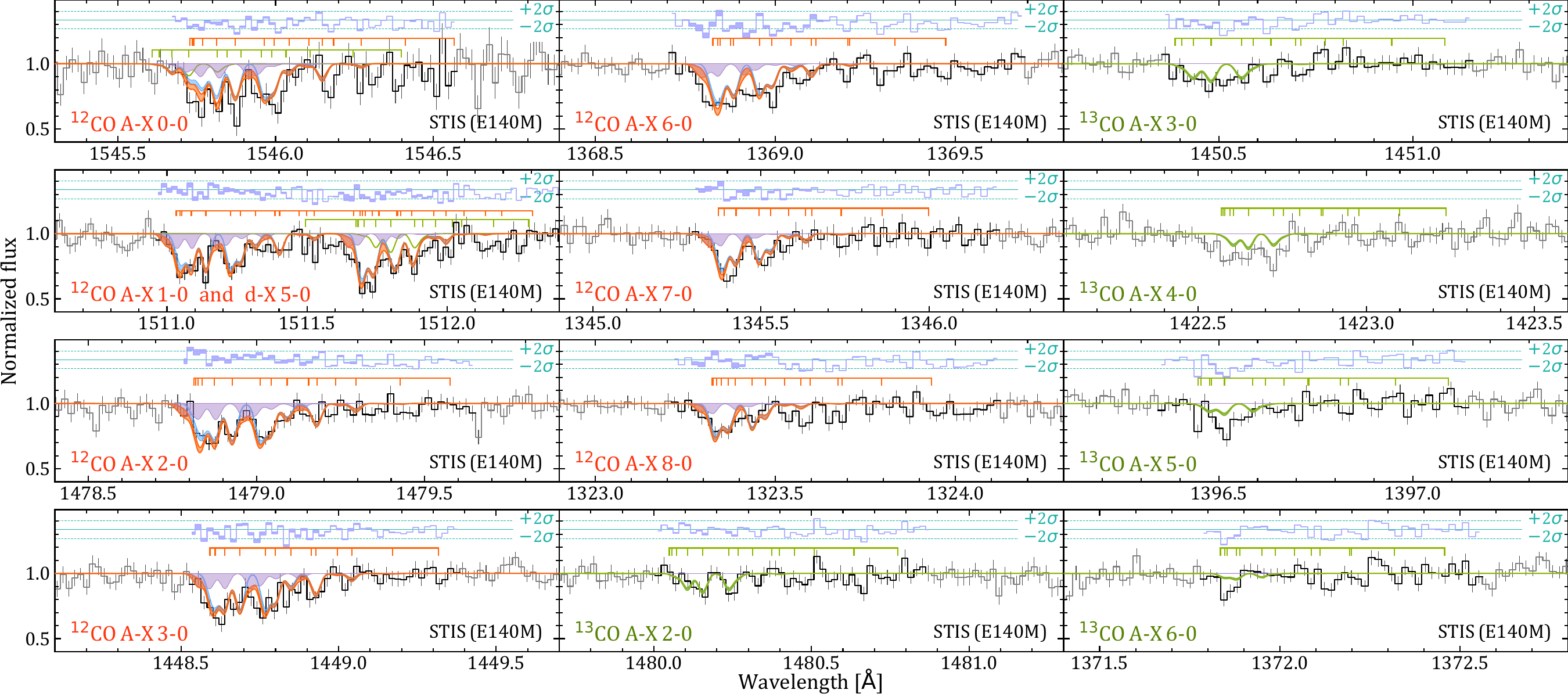}}
  \caption{CO absorption bands in the LMC towards \skss. The instrument (grating) are indicated in the bottom right of each panel. Pixels used to constrain the Voigt-profile model are shown in black, otherwise in grey. The red, blue (green for $^{13}$CO) and violet lines and shaded regions show the model profiles sampled from posterior distribution of fit parameters for the total, main and additional (where CO constrained as an upper limit) component, respectively. The blue lines at the top of each panel shows the residuals between the observed and modeled spectra. The connected ticks mark lines from J=0-5 rotational levels. 
  }
     \label{fig:Sk67_2}
\end{figure*}

\begin{figure*}[!ht]
\centering
\resizebox{1.0\hsize}{!}
        {\includegraphics{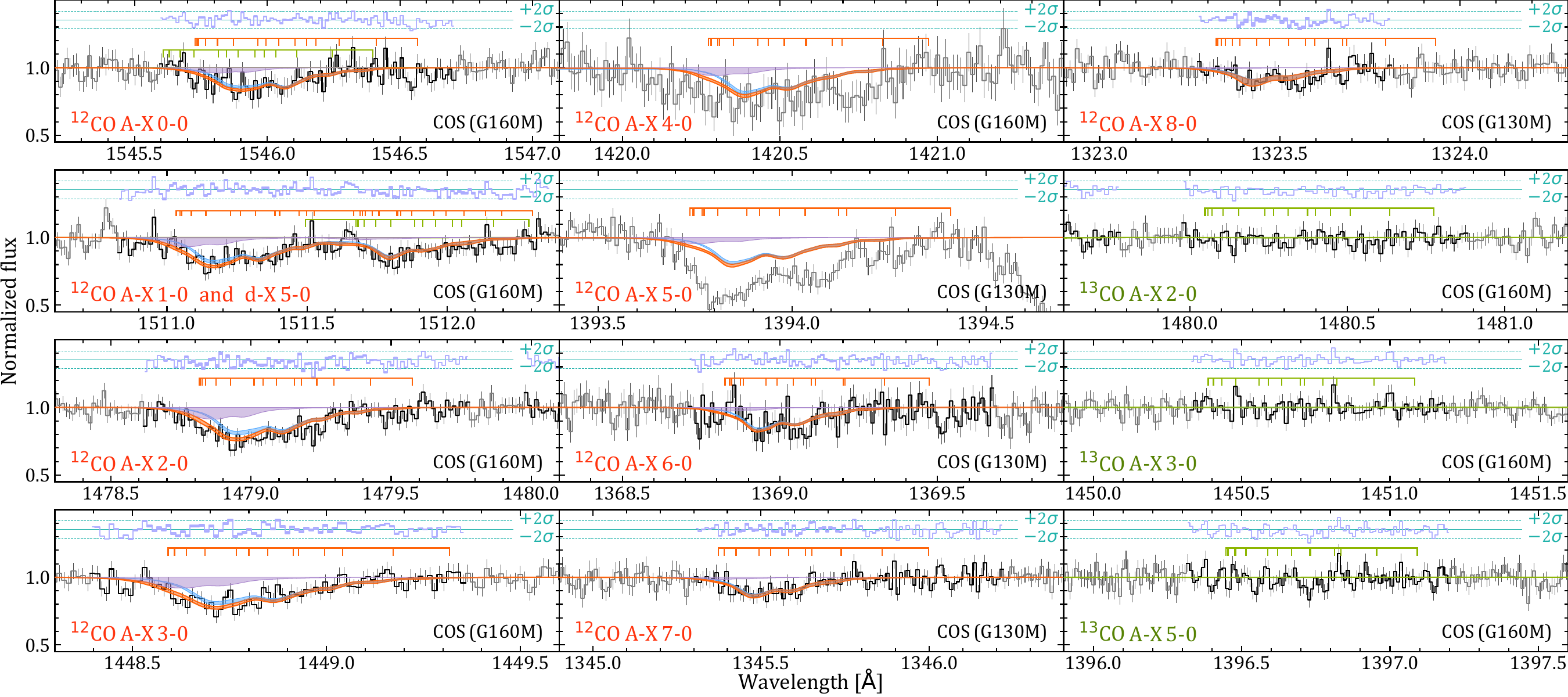}}
  \caption{CO absorption bands in the LMC towards \skse. Graphical elements are the same as in Fig.~\ref{fig:Sk67_2}. The absorption at 1394\,\AA, blended with the $^{12}\rm\,CO\,A-X 5-0$ band, corresponds to the Si\,IV doublet.
          }
     \label{fig:Sk68_137}
\end{figure*}

\begin{figure*}[!ht]
\centering
\resizebox{1.0\hsize}{!}
        {\includegraphics{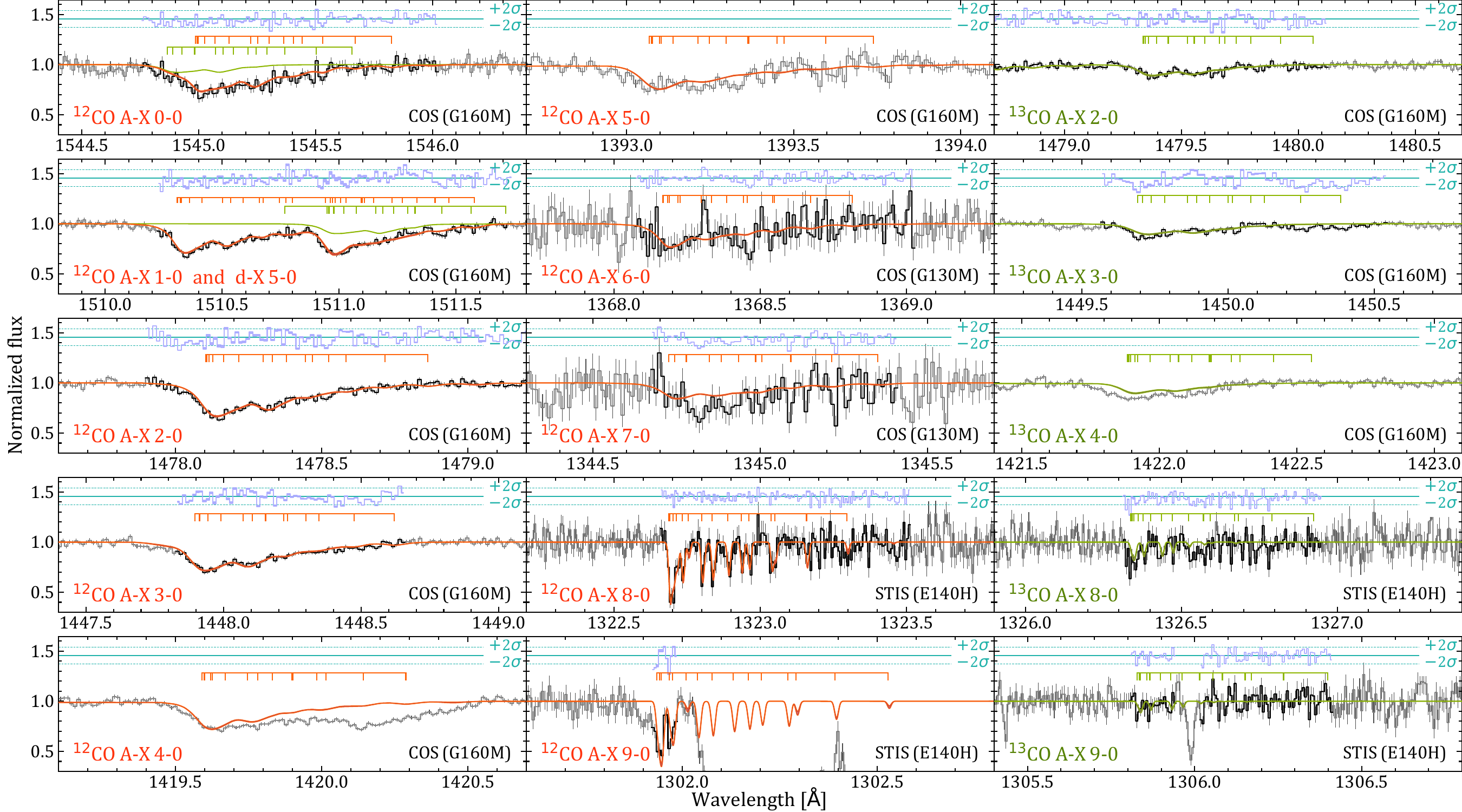}}
  \caption{CO absorption bands in the SMC towards \skoft. Graphical elements are the same as in Fig.~\ref{fig:Sk67_2}. The absorption at $\sim$1420.2\,\AA is due to CO\,$a'-X$\,14-0 band, we did not fit this band.
          }
     \label{fig:Sk143}
\end{figure*}

We confidently detect $^{12}$CO absorption lines towards \skse\ in LMC and both $^{12}$CO and $^{13}$CO towards \skss\ (LMC) and \skoft\ (SMC), yielding an isotopic ratio $\sim 10^{-2}$ consistent with what is seen in the MW (\citealt{Sonnentrucker2007}, \citealt{Sheffer2007}). The fits are shown in Figs.~\ref{fig:Sk67_2}, \ref{fig:Sk68_137} and \ref{fig:Sk143} and the corresponding parameters in Table~\ref{tab:detections}. We also derived stringent upper limits on the $^{12}$CO column densities for 16 and 15 sightlines in the LMC and SMC, respectively. The coadded band profiles are shown in Fig.~\ref{fig:stack_STIS} and the 2$\sigma$ (95.4\% significance level) upper limits are presented in Table~\ref{tab:results}.
Our results, based on several A-X bands observed at high spectral resolution do not confirm previous claims of CO detection from a single C-X band at lower resolution by \citet{Andre2004} and \citet{Bluhm2001}. Our upper limits ($\log N({\rm CO}) <$\,12.9, $<$\,13.2 and $<$\,13.0) are significantly below the reported values of $13.88^{+0.08}_{-0.09}$, $13.77^{+0.20}_{-0.28}$, $13.57^{+0.08}_{-0.09}$ for Sk-67~5, Sk-68 135, Sk-69 246, respectively \citep{Andre2004}\footnote{\citet{Bluhm2001} reported $\log N({\rm CO}) = 13.0\pm 0.4$ towards Sk-69 246.}. 

The best models for the detection are obtained for number densities well above $10^2$ cm$^{-3}$, consistent with the expected values for cold molecular gas. The thermal pressures along the three lines of sight range from $10^4$ to $10^6$ cm$^{-3}$\,K, which is higher than typical values for the cold ISM in the Milky Way \citep{Jenkins2011, Klimenko2020, Klimenko2024}. This elevated pressure is expected, as the Magellanic Clouds have lower metallicity and stronger UV fields than the Milky Way, leading to higher pressures in thermal equilibrium \citep[e.g.][]{Wolfire1995}. Other studies have indeed also reported a trend of higher pressures in the diffuse medium of the Magellanic Clouds compared to that in the Milky Way \citep[see e.g.][]{Welty2016, Kosenko2024}.

The CO absorbing gas exhibits small Doppler parameters, around $0.2-0.5$ km\,s$^{-1}$, and the derived temperatures are below 40~K. While these results may be influenced by the choice of priors (motivated by modeling, see Appendix~\ref{sect:appendix_temp}), in the case of Sk\,143, we found an even lower temperature of $\sim10$\,K in which case the thermal broadening is less than 0.15 km\,s$^{-1}$. This indicates Mach numbers $M \equiv b_{\rm turb}/b_{\rm th}=1.2^{+0.2}_{-0.2}$, $1.7^{+0.5}_{-0.4}$ and $0.7^{+0.1}_{-0.1}$ for \skss, \skse\ and \skoft, respectively, which are consistent with typical values measured in CNM and molecular clouds in our Galaxy \citep[e.g.][]{Heiles2003,Schneider2013}. We also note that the characteristic cloud sizes, 0.01-1\,pc, derived from the column and number densities, are consistent with the \citeauthor{Larson1981} relation, given the observed turbulent motions.

\begin{figure}[!ht]
    \centering
    \begin{tabular}{l}
    \includegraphics[angle=0,width=0.95\columnwidth]{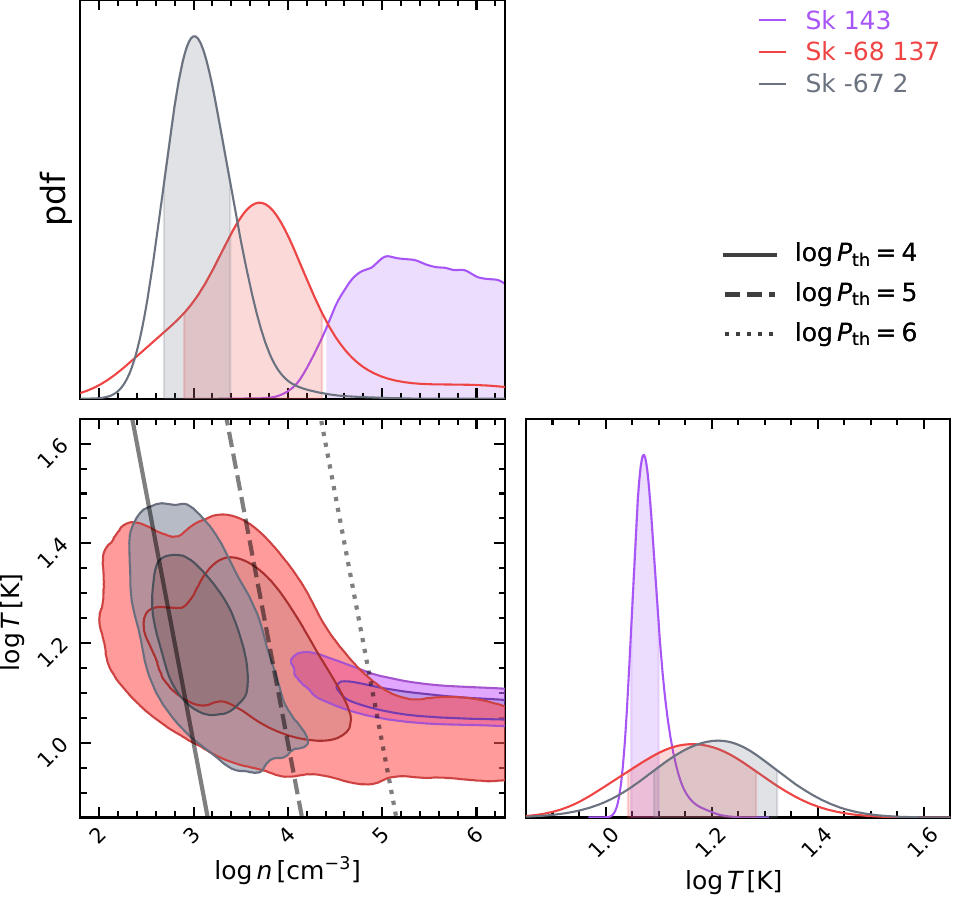} \\  \end{tabular}
    \caption{Marginalized (1D -- diagonal panel, and 2D -- bottom left panel) posterior distributions of the number density and kinetic temperature in CO-bearing gas towards 
    \skss\ (grey), \skse\ (red), and \skoft\ (purple). Shaded regions depict 0.683 (1D and 2D) and 
    0.954 (2D) confidence intervals. Solid, dashed and dotted lines represent constant thermal pressures of $10^{4}$, $10^{5}$ and $10^{6}$\,cm$^{-3}$\,K, respectively. 
    }
    \label{fig:post}
\end{figure}

\renewcommand{\arraystretch}{1.3}
\setlength{\tabcolsep}{1pt}
\begin{table}
\centering
\caption{Results for the CO detections in the Magellanic Clouds 
\label{tab:detections}}
\begin{tabular}{lccccc}
\hline\hline
& \multicolumn{4}{c}{LMC} & SMC \\
Star &  \multicolumn{2}{c}{\skss}  & \multicolumn{2}{c}{\skse} & \skoft \\
\hline
$v_{\rm LSR}^{\dagger}$ (\kms) &  265.1 & 254.3 & 280.9 &  255.3 & 124.3  \\
$b$ (\kms) & $0.40^{+0.05}_{-0.04}$ & $<0.3$ & $0.52^{+0.10}_{-0.10}$ & $1.0^{+0.3}_{-0.3}$ &  $0.20^{+0.01}_{-0.01}$ \\
$\log N(^{12}{\rm CO})$ & $15.42^{+0.35}_{-0.28}$ & $<15.1$ & $14.88^{+0.33}_{-0.19}$ & $<13.7$ &  $16.83^{+0.02}_{-0.02}$ \\
$\log n$  [cm$^{-3}$] & $3.0^{+0.4}_{-0.3}$ & -- & $3.7^{+0.6}_{-0.6}$ & -- & $>4.4^{\ddagger}$  \\
$\log T$  [K] &  $1.2^{+0.1}_{-0.1}$ &  -- & $1.2^{+0.1}_{-0.1}$ &  -- & $1.07^{+0.03}_{-0.02}$ \\
{\Large \strut}$\log {{^{13}{\rm CO}} \over {^{12}{\rm CO}}}$ & $-1.50^{+0.09}_{-0.12}$ & & $<-2.4$ & & $-2.20^{+0.06}_{-0.07}$\\ 
\hline
\end{tabular}
\tablefoot{$\dagger$ Velocity in Local Standard of Rest. 
$\ddagger$\,The  profile becomes insensitive to the density as it exceeds the critical value. 
}
\end{table}

\section{Discussions}
\label{sec:discussion}

\begin{figure}
    \includegraphics[angle=0,width=\columnwidth]{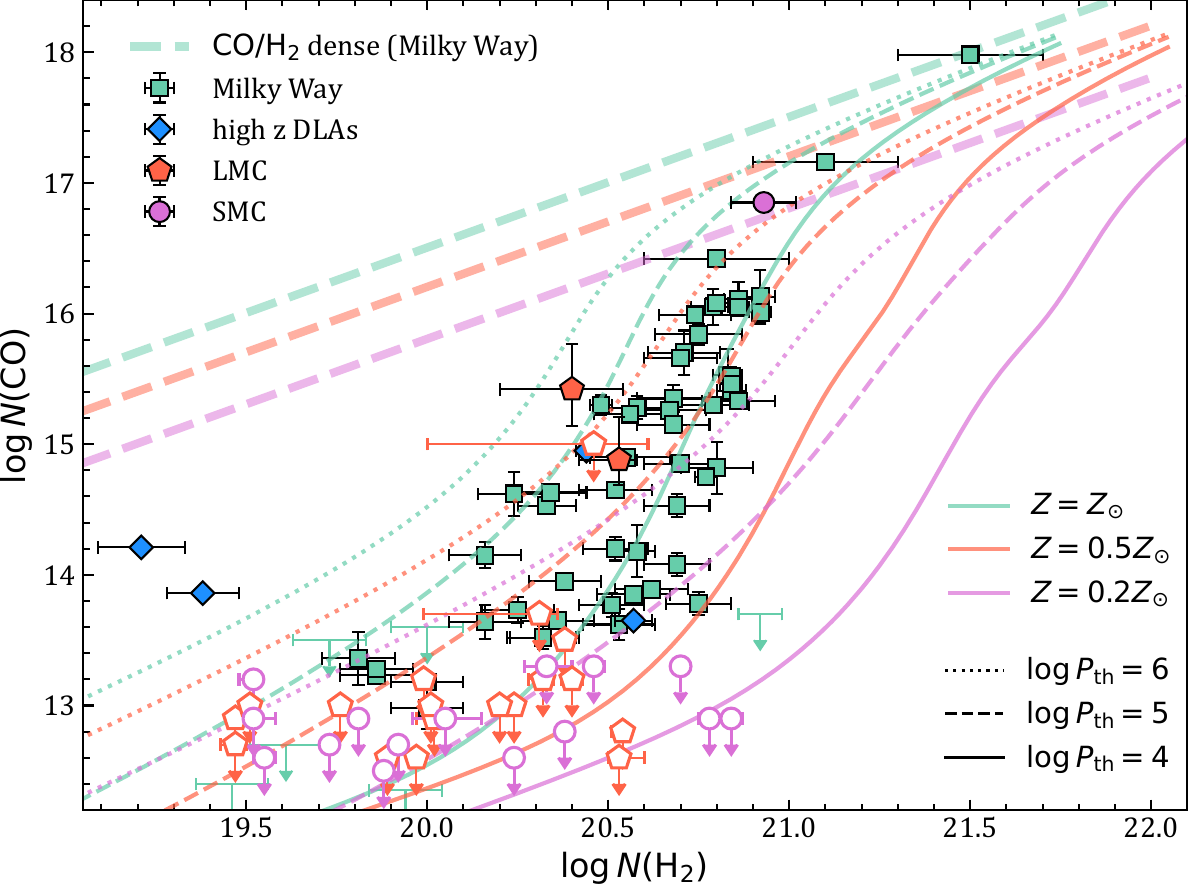}
  \caption{Measurements of CO versus H$_2$ column densities in absorption. Green squares and blue diamonds represent the literature measurements in our Galaxy and high-z DLAs, respectively. The red pentagons and pink circles show the data obtained in this paper for LMC and SMC, respectively. Curves indicate isobaric models using the \texttt{Meudon PDR} code \citep{LePetit2006}, with thermal pressures $10^4$ (solid), $10^5$ (dashed), $10^6$\,cm$^{-3}$\,K (dotted) and metallicities $Z = 0.2$ (pink), 0.5 (red), 1 (green) relative to solar. The dashed lines indicate the standard CO/H$_2=3.2\times10^{-4}$ measured in the dense clouds in the Milky-Way (green) and the same value scaled to the MC's average metallicity (red and purple).
  }
  \label{fig:CO_H2}
\end{figure}

In Fig.~\ref{fig:CO_H2}, we compare our constraints on the $^{12}$CO (hereafter just noted CO) and H$_2$ column densities in the Magellanic Clouds (MCs) with similar absorption measurements in the Milky Way \citep{Sonnentrucker2007, Sheffer2008, Burgh2010, Welty2020, Federman2021} and at high redshifts \citep[][and references therein]{Noterdaeme2018, Klimenko2024}. Most constraints are significantly below the standard ratio CO/H$_2 \approx3\times 10^{-4}$ adopted for the Milky Way \citep{Bolatto2013}. However, they present a steep increase in CO column densities around $\log N(\rm H_2) \sim 20.5$ \cite[see also ][]{Sonnentrucker2007,Sheffer2008}, indicating a transition from atomic to molecular forms of carbon. This also means that majority of the clouds have not reached full carbon molecularization. Remarkably, some SMC and LMC sightlines feature very low CO column densities ($N(\rm CO) \lesssim 13$) even at $\log N(\rm H_2) \gtrsim 20.5$ while two high-$z$ systems present relatively high CO column densities at $\log N(\rm H_2) < 19.5$. The high-$z$ detections can be explained by the Solar or super-Solar metallicities in these particular cases 
\citep{Srianand2008, Noterdaeme2017}. In turn, the MCs have sub-solar metallicities. Environmental differences can also be at play, with e.g. lower pressures in the clouds with low $N$(CO) despite  high $N({\rm H}_2)$. 

The star \skse\ lies north of the 30 Doradus complex, suggesting the absorption system may be part of the star-forming region, so that  the detection of CO may in principle not be surprising. However, CO was not detected in absorption towards other stars near 30 Dor, such as Sk$-$68\,135, Sk$-$69\,246, and Sk$-$68\,129. In contrast, \skss\ and \skoft\, where CO was detected in absorption, are located in more quiescent regions: \skss\ is an isolated star in the northwest of the LMC, and \skoft\ is situated in the eastern part of the SMC wing. Interestingly, \citet{Welty2013} found an unusual chemical composition in the absorption system towards the latter star, differing from typical SMC wing abundances, including the first detection of C$_2$ and C$_3$ outside our galaxy. Overall, the location of the star does not provide a direct hint on the local conditions and hence the detectability of CO.

To quantitatively explore the variation in CO/H$_2$ abundances with physical parameters, we then compare our observations with predictions from the \texttt{Meudon PDR} code \citep{LePetit2006}\footnote{Version 1.5.4, revision 2095 (August 2021)}. Isobaric models were computed for three metallicities: 0.2 and 0.5 (typical for the SMC and LMC, respectively) and 1.0, with respect to Solar. We assumed a plane-parallel geometry with standard dust composition, a UV field intensity equal to the \citealt{Mathis1983} field, and a cosmic ray ionization rate (CRIR) of $10^{-16}$\,s$^{-1}$ per H$_2$ molecule.

Fig.~\ref{fig:CO_H2} shows that, when the metallicity is taken into account, both the MW data and the upper limits in the MCs can be explained by models with thermal pressures of $10^4 - 10^5$\,cm$^{-3}$\,K. However, the CO detections in the LMC and SMC suggest locally either a higher metallicity (closer to Solar) or higher thermal pressures (up to a few $\times 10^6$ cm$^{-3}$K). While large spatial variation of metallicities are seen in the MCs \citep[][]{Tchernyshyov2015, Kosenko2024}, similar to variations seen in the Solar neighbourhood, \citep[][but see \citealt{Ritchey2023}]{DeCia2021}, those along the \skss\ and \skoft\ sightlines are found to be close to the average values \citep[see][]{Kosenko2024,Jenkins2017}. The saturated S\,{\sc ii} lines in \skse\ indicate an elevated metallicity in comparison with the average LMC value. In short, high thermal pressures ($P_{\rm th} > 10^6\,\rm K\,cm^{-3}$) appear to be essential to explain at least two CO detections. 

This is particularly clear for the gas towards \skoft\ in the SMC: The CO/H$_2$ is remarkably high and reaches the maximum value predicted by the corresponding models, that otherwise converge only at much higher $N($H$_2)$ for pressures below that value. In fact, the rotational excitation of CO in this system also indicates a high pressure, as can be seen in Fig.~\ref{fig:post}. In that case, the gas has likely reached full molecularisation so that the CO/H$_2$ ratio, $8.3^{+2.0}_{-1.6} \times 10^{-5}$, should be representative of that of dense molecular gas. We note that exact value of the thermal pressure is hard to estimate since various factors that influence the CO/H$_2$ ratio (metallicity, UV field, { CRIR, and C/H gas abundance}) are not well known in this system.  Interestingly, the ratio is found to be approximately 4 times lower than the standard Galactic value of $3.2\times10^{-4}$ \citep{Bolatto2013}, in agreement with the approximately five times lower metallicity. {However, note that in SMC some studies reported slightly lower carbon abundances $\approx0.1$ of solar value\citep{Welty2016, Vink2023}.} 

In summary, the detection of CO towards \skoft\ enabled the first direct measurement of the CO-to-H$_2$ ratio in the SMC, yielding a value consistent with the standard value, scaled down to the observed SMC metallicity. We caution however that on other factors such as strength of the UV field \citep{Bolatto2013} and CRIR \citep{Bisbas2017} can in principle also alter the ratio, warranting detailed modeling. These could be aided by constraints from \CI\ and H$_2$ abundances and their excitation levels. 

We emphasize that it is essential to ensure a cloud is fully molecularized before applying the observed CO/H$_2$ ratio to any CO-emitting gas. Since very high H$_2$ columns are rare and difficult to characterize in absorption, direct measurements of CO/H$_2$ ratio at low metallicities require observing high-pressure gas. In this context, resolving the CO bands to study the population of rotational levels is crucial.

\begin{acknowledgements}
	  We thank the referee for a very detailed and thorough report, which allowed us to significantly improve the quality of the manuscript. 
      This work is supported by RSF grant 23-12-00166. 
      Based on observations obtained with the NASA/ESA Hubble Space Telescope, retrieved from the Mikulski Archive for Space Telescopes (MAST) at the Space Telescope Science Institute (STScI). STScI is operated by the Association of Universities for Research in Astronomy, Inc. under NASA contract NAS 5-26555.
\end{acknowledgements}

\bibliographystyle{aa}


\appendix
\section{Model of CO excitation}
\label{sect:CO_model}
Here we describe the CO excitation model used to tie column densities across the various rotational levels during Voigt profile fitting. We employed a one-zone static model, where the balance equations for each $i$th level are expressed as
\begin{equation}
\sum_{j} f_j Q_{ij} = f_i\sum_{j}Q_{ji},
\end{equation}
where $f_i = n_{i} / n_{\rm tot} = N_{i} / N_{\rm tot}$, and $n_i$\,($N_{i}$) and $n_{\rm tot}$\,($N_{\rm tot}$) represent the CO number (column) density in the $i$th level and total, respectively.
$Q_{ij}$ are the transition rates between $i$th and $j$th levels. In our model, we accounted for populating the levels by collisions and CMB radiation, and also incorporated the radiative trapping effect:

\begin{equation}
Q_{ul} = \beta \times A_{ul} + \beta\times I_{\rm CMB} B_{ul} + \sum_{k}n^k C^k_{ul},
\end{equation} 
\begin{equation}
Q_{lu} = \beta\times I_{\rm CMB} B_{lu} + \sum_kn^k C^k_{lu}
\end{equation}
where $u$ and $l$ correspond to the upper and lower levels, respectively. The  terms in both $Q_{ul}$ and $Q_{lu}$ describe collisional excitation: $n^k$ is the number density of collisional partner $k$, $C_{ul}^k$ and $C_{lu}^k$ are the corresponding collisional rate coefficients. Assuming that CO resides in a H$_2$-dominated medium, we considered collisions with H$_2$ (with ortho-para ratio of H$_2$ corresponding to the kinetic temperature) and He (with 7\% abundance) with collisional rates taken from \citealt{Yang2010,Cecchi-Pestellini2002}. $A_{ul}$, $B_{lu}$ and $B_{ul}$ are Einstein coefficients \citep[taken from][]{Schoier2005}, describing spontaneous decay, absorption and stimulated emission. $I_{\rm CMB}$ is the intensity of CMB radiation at $z=0$, with $T_{\rm CMB}=2.725$\,K. $\beta \sim N_{\rm tot}({\rm CO})$ is the escape probability coefficient, which accounts for radiative trapping effect \citep[see][]{Klimenko2024}. Radiative trapping begins to play a significant role for column densities $\log N_{\rm tot}({\rm CO}) \gtrsim 15$, 
where we assume an exponential dependence of $\beta$ on the total CO column density, $\beta = e^{- (\log N_{\rm tot}({\rm CO}) - 15) / 2}$, based on extrapolating points obtained through detailed calculations \citep{Klimenko2024}. We note  that for two detections with relatively large CO column densities, the joint line profiles of CO bands are better reproduced by collisional excitation, making radiative trapping negligible in these cases. Overall, the described model provides column densities for the various rotational levels as a function of $N_{\rm tot}({\rm CO})$, $n$ and $T_{\rm K}$, which were used as independent parameters during the line profile fitting.

\section{Model motivated choice of temperature prior}
\label{sect:appendix_temp}
\begin{figure}[!h]
\centering
    \includegraphics[angle=0,width=\columnwidth]{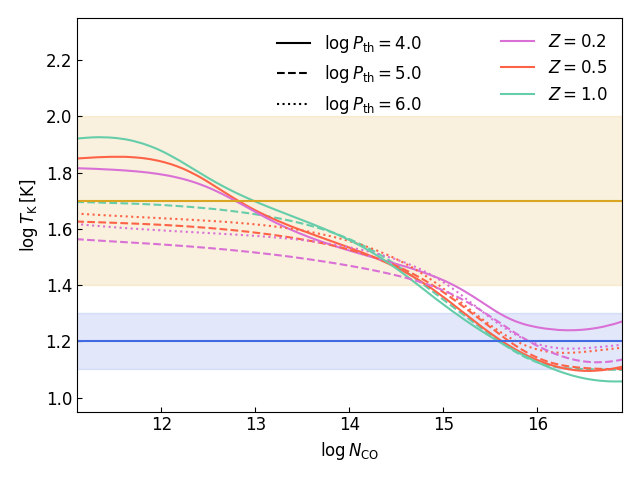} 
    \caption{Kinetic temperature as a function of CO column density obtained using the \texttt{Meudon PDR} code. Solid, dashed and dotted lines depict  models with thermal pressures $10^4$, $10^5$ and $10^6$ cm$^{-3}$\,K, respectively, while pink, red and green colors correspond to metallicities (relative to solar) $Z = 0.2$, 0.5 and 1, respectively. 
    The blue ($\log T_{\rm K}[\rm K] = 1.2\pm 0.1$) and orange ($\log T_{\rm K}[\rm K] = 1.7\pm 0.3$) horizontal stripes show our choice of priors on the kinetic temperatures used to fit CO absorption lines towards sightlines with and without CO detections, respectively. 
    }
    \label{fig:temp_priors}
\end{figure}

In Fig.~\ref{fig:temp_priors} we compare the kinetic temperature profiles as a function of $N(\rm CO)$, for a set of Meudon PDR models with different metallicities and thermal pressures (these models are described and used in  Sect.~\ref{sec:discussion}). One can notice that models present a similar behaviour in the CO column density range $\log N(\rm CO) = 13-16$, with median $T\approx50$\,K for $\log N(\rm CO) = 13$ (similar to detection limits) and $T\approx15$\,K for $\log N(\rm CO) = 16$. These motivated our choice of priors on the kinetic temperature (see Section~\ref{sec:data}) during the fit for both non- detections and detections, respectively.

\section{Overall sample: non-detections and summary}
\label{sect:appendix}

\begin{figure*}
\begin{tabular}{cc}
    {\includegraphics[trim=0 0cm 0 0, angle=0,width=\columnwidth]{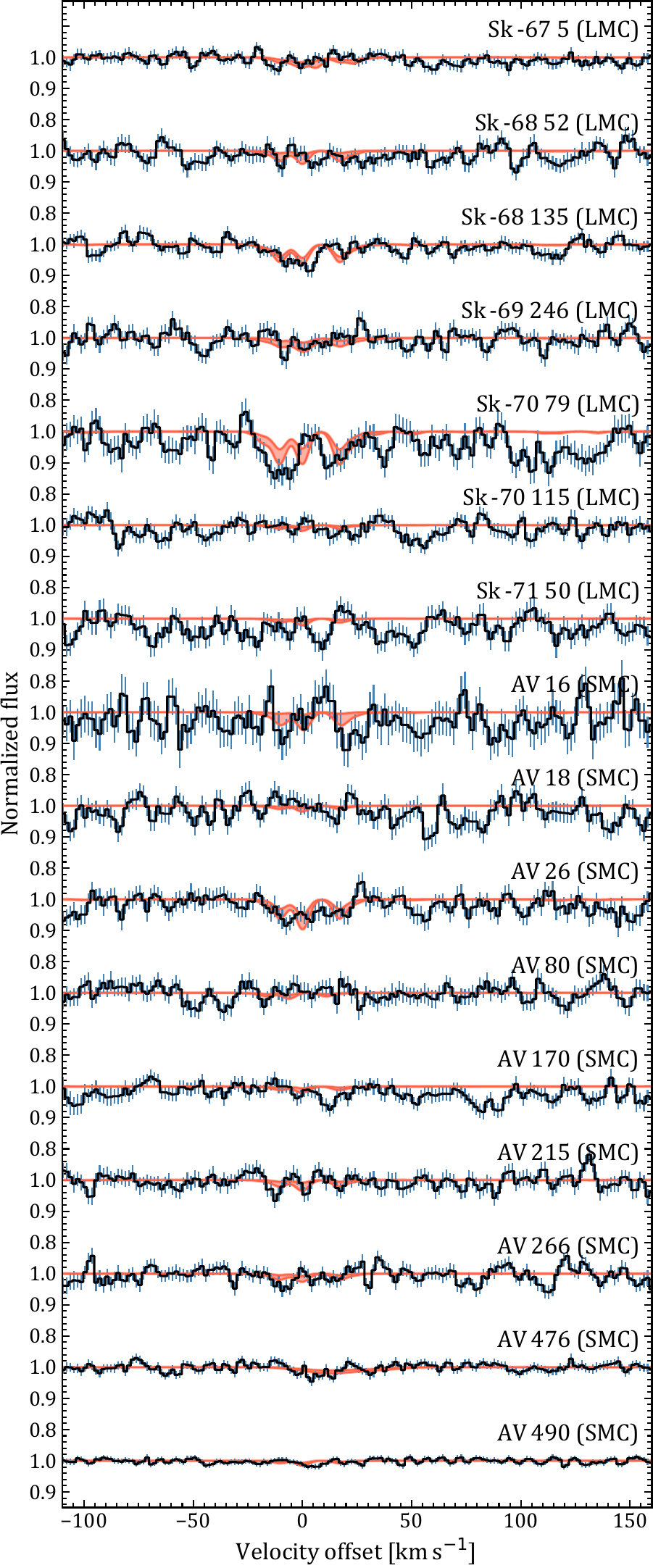}} &
    {\includegraphics[trim=0 0cm 0 0, angle=0,width=\columnwidth]{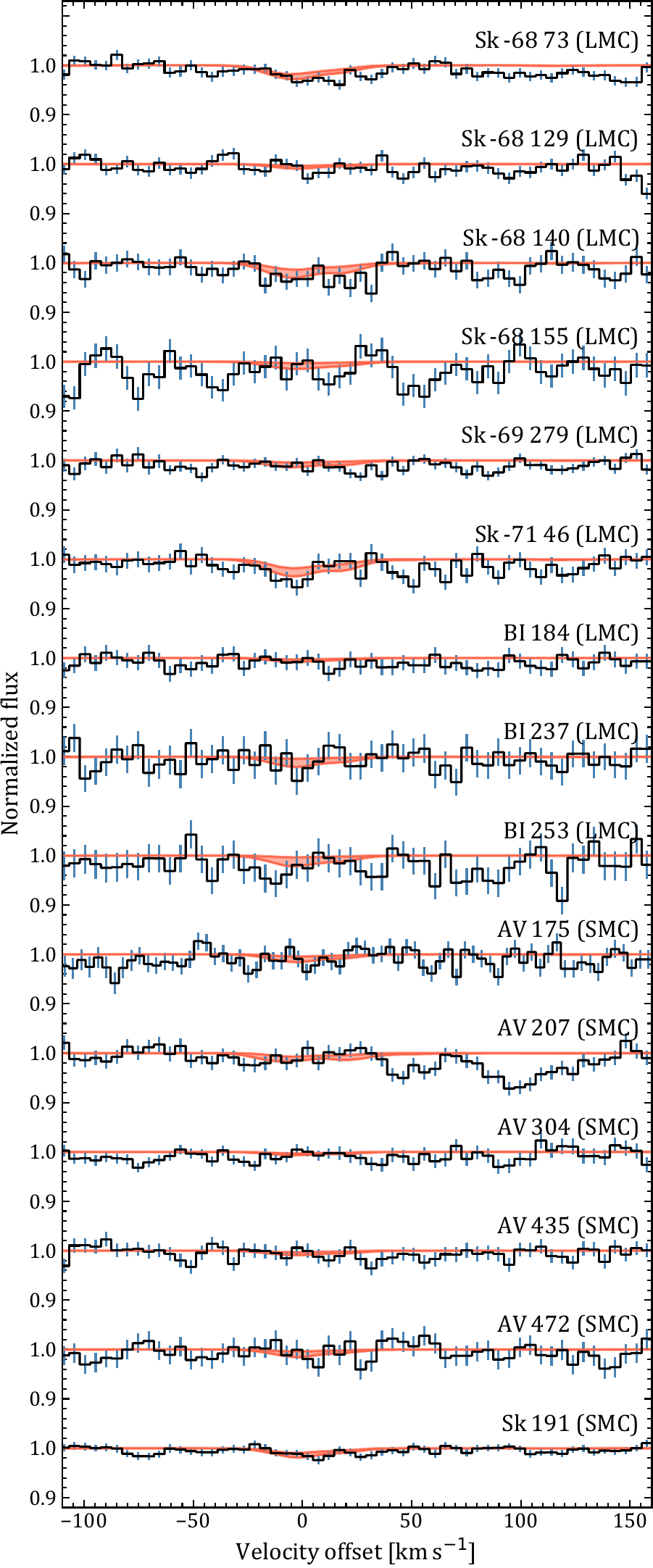}} \\
    \end{tabular}
    \caption{Stack of STIS (left column) and COS (right column) HST spectra of CO absorption lines towards LMC and SMC stars. The black lines with blue errorbars represent the CO bands co-added spectrum (using R0 line as a reference line for each band to define the velocity offset), while the red lines correspond to the co-added fit profiles. Each spectrum is arbitrary shifted in y-axis for illustrative purposes. The name of each background star is provided above each spectrum. This includes three sightlines Sk\,$-$67 5, Sk\,$-$68 135 and Sk\,$-$69 246, where  \citet{Bluhm2001,Andre2004} previously claimed CO detection, while HST data indicate non-detection with much lower upper limits on CO column densities, see Table~\ref{tab:results}.}
     \label{fig:stack_STIS}
\end{figure*}

Figs.~\ref{fig:stack_STIS} presents the coadded spectra at the position of CO absorption bands for the sightlines without CO detection. To coadd the spectra and fit model we shifted the lines in each band using the R0 line as reference, determining the velocity offset (Fig.~\ref{fig:stack_STIS}). This coadding was performed for illustrative purposes only; constraints on the column densities were obtained using the full spectral information. This technique is also described in \citep[][]{Noterdaeme2018}. Some of the sightlines, e.g. Sk\,$-$68\,135 or Sk\,$-$70\,79, present tentative detections only, with constraints on CO column densities close to the expected values for the typical physical conditions in the diffuse ISM. However, deeper and higher resolution studies are needed to substantiate these possible detections.

Table~\ref{tab:results} summaries all the measurements and upper limits obtained towards the 34 stars in the LMC and SMC, together with some additional information from the literature. 
\renewcommand{\arraystretch}{1.25}
\begin{table*}
\caption{Summary of CO and H$_2$ measurements towards LMC and SMC sightlines. } 
\label{tab:results}      
\centering          
\begin{tabular}{l c c c c c c c}
\hline\hline       
Star & $\log N(\rm HI)^{a}$ & $v_{\rm LSR}$ & $\log N(\rm H_2)^{b}$ & $\log N(\rm CO)^{c}$ & $\log N(\rm CI)^d$ & instrument & comment \\ 
 & $[\rm cm^{-2}]$ & km\,s$^{-1}$ & $[\rm cm^{-2}]$ & $[\rm cm^{-2}]$ & $[\rm cm^{-2}]$ & & \\
\hline             
\multicolumn{8}{c}{\sl Large Magellanic Cloud}\\
   Sk\,$-$67\,2 & 21.00 & 254 & $20.46^{+0.15}_{-0.46}$ & $<15.0$ & $13.89^{+0.02}_{-0.09}$ & STIS & \\  
                &  &  263 & $20.40^{+0.14}_{-0.20}$ & $15.42^{+0.35}_{-0.28}$ & $14.32^{+0.04}_{-0.04}$ &  STIS & CH, CH$^{+}$, {CN} \citep{Welty2006}\\
   Sk\,$-$68\,137 & 21.50 & 255 & $20.31^{+0.05}_{-0.32}$ & $<13.7$ & $14.29^{+0.02}_{-0.02}$$^c$ & COS &  \\  
                &  & 281 & $20.53^{+0.05}_{-0.11}$ & $14.88^{+0.33}_{-0.19}$ & $16.2^{+0.6}_{-0.3}$$^{c,e}$ &  COS & \\
   Sk\,$-$67\,5 & 21.00 & 278 & $19.47^{+0.01}_{-0.01}$ & $<12.9$ & $14.02^{+0.02}_{-0.02}$ & STIS & \citet{Andre2004} \\
   Sk\,$-$68\,135 & 21.46 & 262 & $19.99^{+0.01}_{-0.01}$ & $<13.2$ & $14.43^{+0.04}_{-0.03}$ &  STIS & \citet{Andre2004}, CH, CH$^+$ \citep{Welty2006} \\
   Sk\,$-$69\,246 & 21.47 & 267 & $19.76^{+0.01}_{-0.01}$ & $<13.0$ & $14.18^{+0.04}_{-0.03}$ & STIS & \citet{Bluhm2001, Andre2004}, \\
   & & & & & & & CH, CH$^+$ \citep{Welty2006}  \\
   Sk\,$-$68\,52  & 21.30 & 229 & $19.51^{+0.01}_{-0.01}$ & $<13.0$ & $14.07^{+0.04}_{-0.04}$ &  STIS & CH, CH$^+$ \citep{Welty2006} \\
   Sk\,$-$68\,73 & 21.66 & 279 & $20.24^{+0.01}_{-0.01}$ & $<13.0$ & $14.59^{+0.06}_{-0.03}$ & COS+STIS & CH, CH$^{+}$ \citep{Welty2006} \\
   Sk\,$-$68\,129 & 21.72 & 259 & $20.53^{+0.07}_{-0.03}$ & $<12.6$ & $14.05^{+0.01}_{-0.01}$ & COS &  \\
   Sk\,$-$68\,140 & 21.47 & 261 & $20.40^{+0.03}_{-0.03}$ & $<13.2$ & $14.12^{+0.01}_{-0.01}$ & COS &  \\
   Sk\,$-$68\,155 & 21.44 & 279 & $20.02^{+0.03}_{-0.05}$ & $<12.9$ & $14.00^{+0.01}_{-0.01}$ & COS &  \\
   Sk\,$-$69\,279 & 21.59 & 262 & $20.54^{+0.01}_{-0.03}$ & $<12.8$ & $13.96^{+0.43}_{-0.20}$ & COS & \\
   Sk\,$-$70\,79 & 21.26 & 220 & $20.38^{+0.01}_{-0.01}$ & $<13.5$ & $14.31^{+0.01}_{-0.01}$ & STIS &  \\
   Sk\,$-$70\,115 & 21.13 & 208 & $19.97^{+0.01}_{-0.01}$ & $<12.6$ & $13.91^{+0.01}_{-0.01}$ & STIS(H+M) & CH, CH$^+$ \citep{Welty2006} \\
   Sk\,$-$71\,46 & -- & 227 & $20.32^{+0.03}_{-0.04}$ & $<13.2$ & $14.09^{+0.04}_{-0.04}$ & COS & \\
   Sk\,$-$71\,50 & 21.18 & 262 & $19.47^{+0.03}_{-0.04}$ & $<12.7$ & $13.77^{+0.03}_{-0.03}$ & STIS & \\
   BI\,184 & 21.12 & 239 & $19.89^{+0.01}_{-0.02}$ & $<12.6$ &  $13.92^{+0.04}_{-0.03}$ & COS & \\
   BI\,237 & 21.62 & 276 & $20.20^{+0.01}_{-0.01}$ & $<13.0$ & $13.74^{+0.03}_{-0.02}$ & COS & \\
   BI\,253  & 21.67 & 260 & $20.01^{+0.01}_{-0.01}$ & $<13.0$ & $14.00^{+0.02}_{-0.03}$ & COS & \\
   \hline
    \multicolumn{8}{c}{\sl Small Magellanic Cloud}\\
   Sk\,143 & 21.00 & 124 & $20.93^{+0.09}_{-0.09}$$^f$  & $16.85^{+0.02}_{-0.02}$ & $14.93^{+0.06}_{-0.06}$$^c$ & STIS(H)+COS &  CH, CN, C$_2$, C$_3$ \citep{Welty2013} \\
   AV\,16  & -- & 117 & $20.33^{+0.07}_{-0.06}$ & $<13.3$ & $13.86^{+0.04}_{-0.05}$$^c$ & STIS & \\ 
   AV\,18  & 22.04 & 139 & $20.46^{+0.03}_{-0.02}$ & $<12.7$ &  $13.42^{+0.03}_{-0.03}$$^c$ & STIS & \\
   AV\,26  & 21.70 & 116 & $20.70^{+0.02}_{-0.01}$ & $<13.3$ & $15.15^{+0.12}_{-0.10}$ & STIS(H+M) & CH \citep{Welty2006} \\ 
   AV\,80  & 21.81 & 110 & $20.24^{+0.01}_{-0.01}$ & $<12.6$ & $13.61^{+0.03}_{-0.03}$ & STIS & \\ 
   AV\,170 & 21.14 & 118 & $19.73^{+0.01}_{-0.01}$ & $<12.7$ & $13.41^{+0.06}_{-0.06}$ & STIS & \\ 
   AV\,175 & -- & 137 & $20.05^{+0.10}_{-0.09}$ & $<12.9$ & $13.20^{+0.05}_{-0.09}$ & COS & \\ 
   AV\,207 & 21.43 & 153 & $19.52^{+0.02}_{-0.04}$ & $<13.2$ & $13.59^{+0.03}_{-0.03}$ & COS & \\
   AV\,215 & 21.86 & 123 & $19.52^{+0.06}_{-0.05}$ & $<12.9$ & $13.79^{+0.14}_{-0.12}$ & STIS & \\
   AV\,266 & -- & 119 & $19.81^{+0.01}_{-0.02}$ & $<12.9$ &  $13.48^{+0.12}_{-0.11}$ & STIS &  \\
   AV\,304 & 21.48 & 114 & $19.55^{+0.03}_{-0.03}$ & $<12.6$ & $13.47^{+0.04}_{-0.03}$$^c$ & COS &  \\
   AV\,435 & 21.54 & 176 & $19.92^{+0.01}_{-0.01}$ & $<12.7$ & $13.29^{+0.05}_{-0.04}$$^c$ & COS & \\
   AV\,472 & -- & 119 & $20.38^{+0.01}_{-0.01}$ & $<12.8$ & $13.7^{+0.5}_{-0.4}$$^c$ & COS & \\ 
   AV\,476 & 21.85 & 160 & $20.84^{+0.03}_{-0.05}$ & $<12.9$ & $14.62^{+0.12}_{-0.11}$ & COS & CH, CH$^+$ \citep{Welty2006} \\  
   AV\,490 & 21.46 & 125 & $19.88^{+0.01}_{-0.01}$ & $<12.5$ &  $13.63^{+0.01}_{-0.01}$ & STIS &  \\
   Sk\,191 & 21.51 & 145 & $20.78^{+0.02}_{-0.03}$ & $<12.9$ & $13.77^{+0.03}_{-0.02}$ & STIS+COS & \\    
\hline                  
\end{tabular}

\tablebib{
 \tablefoottext{a}{\cite{Welty2012, Roman_Duval2019}};
 \tablefoottext{b}{\citet[][unless specified otherwise]{Kosenko2023}};
  \tablefoottext{c}{This work};
 \tablefoottext{d}{\citet[][unless specified otherwise]{Kosenko2024}};
 \tablefoottext{e}{The line profiles are saturated in COS spectrum therefore we propose to consider this value with caution}.
 \tablefoottext{f}{{\bf\cite{Cartledge2005}}}; 
}

\end{table*}

\section{CO excitation diagrams}
\begin{figure}[!h]
\centering
    \includegraphics[angle=0,width=\columnwidth]{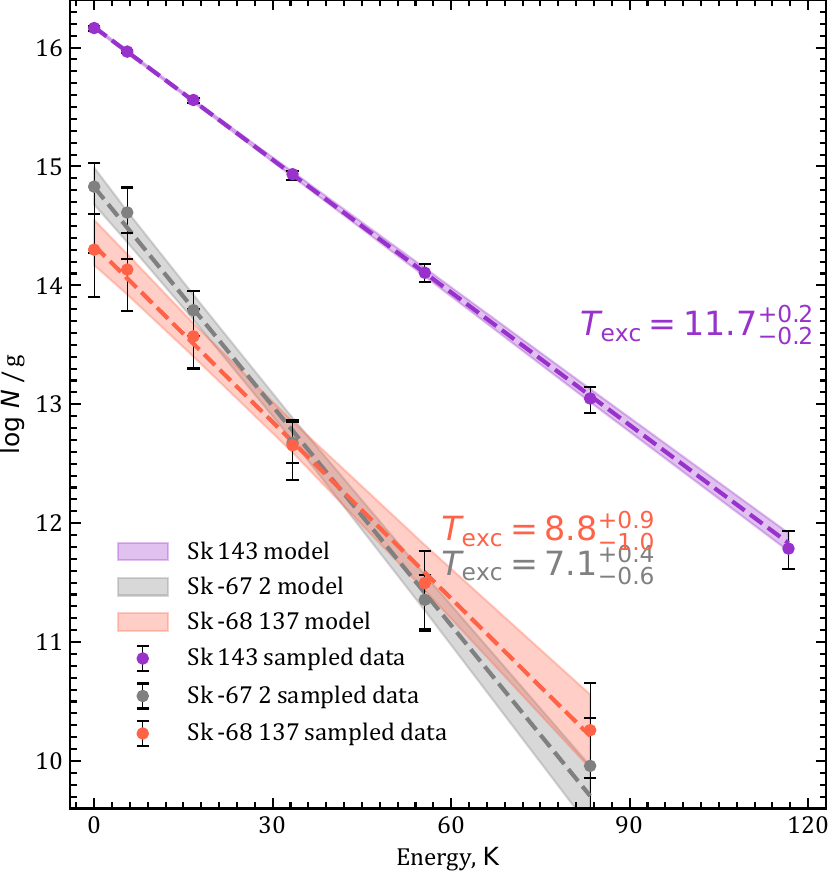} 
    \caption{The CO excitation diagram towards three CO absorption systems. The gray, red and violet points indicate the column densities values sampled from the posterior distributions of the fit parameters using the model described in Section~\ref{sec:data} for \skss, \skse, and \skoft, respectively. The stripes of the corresponding colors indicate the 0.683 credible region of the excitation diagram derived from the fit to the sampled column densities with Boltzmann law.
    \label{fig:exc_diag}
    }
\end{figure}

Using the sampling from derived posterior distribution function for the physical conditions ($n$, $T_{\rm K}$) and the total CO column density, we can reconstruct the CO excitation diagram for {each} detected system. The data is provided in Table~\ref{tab:col_dens}. An example of excitation diagram is shown in Fig.~\ref{fig:exc_diag}. For \skoft\  the fit to the sampled CO column densities with Boltzmann distribution indicate a temperature $T_{\rm exc} = 11.7\pm0.2$\,K. These estimates match the constrained estimates obtained directly from the model fit, which is expected since the posterior values of the number densities are well above the critical densities for all rotational levels. We also note that constrained values are consistent with the trend of increase of the $T_{\rm exc}$ at CO column densities $\log N(\rm CO)\gtrsim 15$ seen in our Galaxy \citep[see e.g. compilations by][]{Sonnentrucker2007,Klimenko2024}.

\setlength{\tabcolsep}{1pt}
\begin{table}
\centering
\caption{The CO rotational column densities 
\label{tab:col_dens}}
\begin{tabular}{lccc}
\hline\hline
rotational & \multicolumn{3}{c}{$\log N (^{12}{\rm CO})$} \\
level &  \skss  & \skse & \skoft \\
\hline
$J=0$  & $14.83^{+0.56}_{-0.20}$ & $14.3^{+0.4}_{-0.3}$ & $16.164^{+0.023}_{-0.018}$ \\
$J=1$  & $15.09^{+0.39}_{-0.21}$ & $14.61^{+0.35}_{-0.31}$ & $16.443^{+0.015}_{-0.015}$ \\
$J=2$  & $14.49^{+0.22}_{-0.16}$ & $14.27^{+0.27}_{-0.23}$ & $16.258^{+0.025}_{-0.020}$ \\
$J=3$  & $13.52^{+0.17}_{-0.18}$ & $13.50^{+0.29}_{-0.21}$ & $15.78^{+0.05}_{-0.03}$ \\
$J=4$  & $12.31^{+0.25}_{-0.21}$ & $12.45^{+0.40}_{-0.27}$ & $15.06^{+0.08}_{-0.07}$ \\
$J=5$  & $11.0^{+0.4}_{-0.4}$ & $11.3^{+0.4}_{-0.4}$ & $14.09^{+0.12}_{-0.10}$ \\
$J=6$  & - & - & $12.90^{+0.17}_{-0.15}$ \\
\hline
$T_{\rm exc},\,K$ & $7.1^{+0.4}_{-0.6}$ & $8.8^{+0.9}_{-1.0}$ & $11.7^{+0.2}_{-0.2}$ \\
\end{tabular}
\end{table}

\end{document}